\documentclass[amsmath,amssymb,amsbsy,prb,twocolumn,preprintnumbers,showpacs]{revtex4-2}
\usepackage{graphicx,color}
\usepackage{dcolumn}
\usepackage{bm}
\usepackage{braket}
\usepackage{mathtools}
\usepackage{ulem}
\usepackage{mathrsfs}
\usepackage{times}
\usepackage[breaklinks,colorlinks=true,linkcolor=blue,urlcolor=blue,citecolor=blue]{hyperref}

\begin{document}
\title{Flat band, spin-1 Dirac cone, and Hofstadter diagram in the fermionic square kagome model}

\author{Tomonari Mizoguchi}
\affiliation{Department of Physics, University of Tsukuba, Tsukuba, Ibaraki 305-8571, Japan}
\email{mizoguchi@rhodia.ph.tsukuba.ac.jp}
\author{Yoshihito Kuno}
\affiliation{Department of Physics, University of Tsukuba, Tsukuba, Ibaraki 305-8571, Japan}
\author{Yasuhiro Hatsugai}
\affiliation{Department of Physics, University of Tsukuba, Tsukuba, Ibaraki 305-8571, Japan}

\date{\today}
\begin{abstract}
We study characteristic band structures of the fermions on a square kagome lattice,
one of the two-dimensional lattices hosting a corner-sharing network of triangles. 
We show that the band structures of the nearest-neighbor tight-binding model 
exhibit many characteristic features, including a flat band which is ubiquitous among frustrated lattices.
On the flat band, we elucidate its origin by using the molecular-orbital representation,
and  also find localized exact eigenstates called compact localized states.
In addition to the flat band, we also find two spin-1 Dirac cones with different energies.
These spin-1 Dirac cones are not described by the simplest effective Dirac Hamiltonian
because the middle band is bended and the energy spectrum is particle-hole asymmetric. 
We also investigated the Hofstadter problem on a square kagome lattice in the presence of an external field, 
and find that the profile of the Chern numbers around the modified spin-1 Dirac cones coincides with the conventional one. 
\end{abstract}

\maketitle
\section{Introduction}
Geometrical frustration is a source of exotic physics in condensed matter systems. 
For instance, antiferromagnetic localized spin systems on geometrically-frustrated lattices  
are candidates of an exotic non-magnetic ground state, called quantum spin liquid~\cite{Balents2010,Zhou2017}.
A triangle is a basic building block that produces frustration.
Therefore, triangle-based lattice structures, such as a kagome lattice (in two dimensions) 
and a pyrochlore lattice (in three dimensions), 
serve as a fertile ground to study the roles of geometrical frustration.

Geometrical frustration also produces interesting features 
in the dispersion relation of tight-binding models~\cite{Mielke1991,Mielke1991_2,Bergman2008,Hatsugai2011,Kariyado2013,Yamashita2014}.
Typically, frustration results in the emergence of a completely dispersionless band~\cite{Mielke1991,Mielke1991_2,Bergman2008,Hatsugai2011}, 
called a flat band,
which is also related to the emergence of spin liquid states of the spin models~\cite{Reimers1991,Garanin1999,Isakov2004}.
The Dirac cones are another feature which is often seen in frustrated tight-binding models~\cite{Kariyado2013,Yamashita2014,Essafi2017}.

\begin{figure}[b]
\begin{center}
\includegraphics[clip, width = 0.85\linewidth]{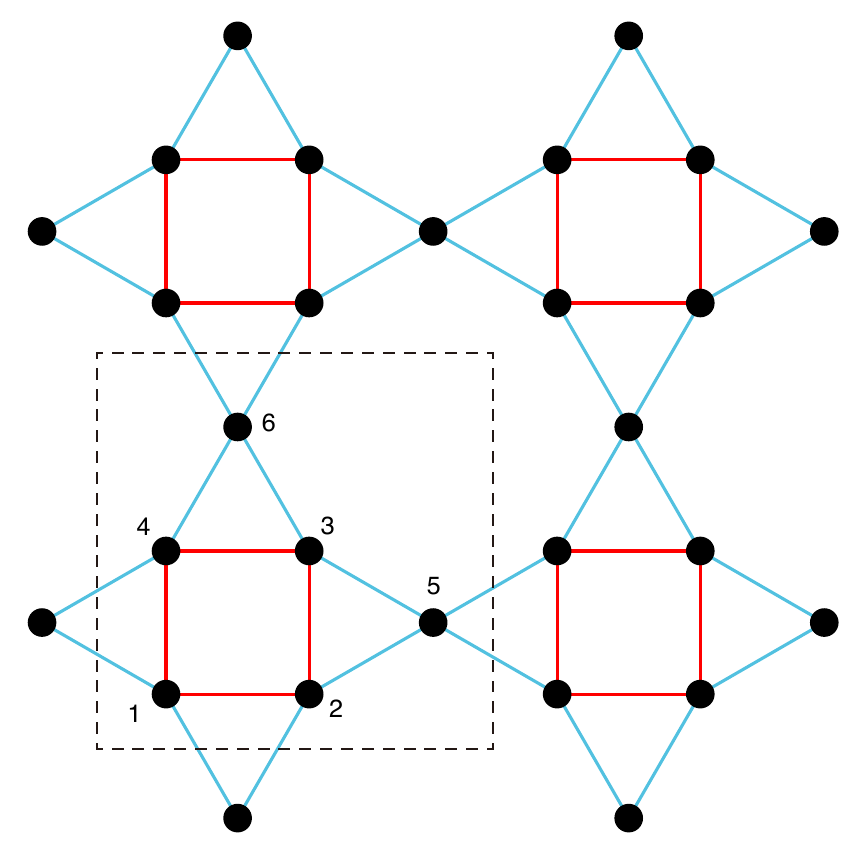}
\vspace{-10pt}
\caption{A square kagome lattice. A black dashed square denotes the unit cell. }
  \label{fig:model}
 \end{center}
 \vspace{-10pt}
\end{figure}

With these as backgrounds, in this paper, we investigate the band structures of 
the nearest-neighbor (NN) tight-binding model on a square kagome lattice (Fig.~\ref{fig:model}),
which is one of the corner-sharing networks of triangles in two dimensions.
Obviously, the lattice structure has a high frustration. 
In fact, the localized spin model on this lattice structure has been studied in the context of frustrated magnetism~\cite{Siddharthan2001,Tomczak2003,Derzhko2006,Richter2004,Wildeboer2011,Nakano2013,Derzhko2013,Rousochatzakis2013,Derzhko2014,Ralko2015,Pohle2016,Morita2018,Hasegawa2018,Lugan2019}.
Furthermore, beyond the purely theoretical interests, 
the lattice was recently found to be relevant to the real material~\cite{Fujihala2020}.
However, the band structures of the fermionic model are less understood compared 
with its cousin, namely the kagome lattice, which motivated us to study the present model in details. 

We find the two characteristic features of the band structures appear in different energies, 
namely, the flat band and the spin-1 Dirac cones.
We first elucidate the origin of the flat band by using the molecular-orbital 
(MO) representation~\cite{Hatsugai2011,Hatsugai2015,Mizoguchi2019,Mizoguchi2019_2,Mizoguchi2020},
namely, we rewrite the Hamiltonian by using the non-orthogonal basis that is composed of few atomic sites.
As a bonus of this rewriting, we can easily find a set of localized states that corresponds to the flat band in the momentum-space representation.
Such states are helpful for considering the intriguing 
many-body effects such as Wigner crystallization~\cite{Wu2007}, 
superconductivity~\cite{Huber2010}, and quantum many body scars~\cite{Kuno2020_scar}. 
As for the spin-1 Dirac cones, which 
is a triple band touching including linearly dispersive bands,
they appear at two different momenta with different energies. 
We find that some interesting features which can not be described by the simple Dirac Hamiltonian. 
This indicates that some symmetry-allowed modifications are needed
 to describe these spin-1 Dirac cones.
To further reveal the effect of topological gap-opening of the modified spin-1 Dirac cones,
we consider the Hofstadter problem in the presence of the external magnetic field.
The Hofstadter problem, which deals with the fermionic lattice models under the magnetic field, 
has been considered on various lattices~\cite{Hofstadter1976,Hatsugai1990,Aoki1996,Kimura2002,Hatsugai2006}, 
and novel aspects of this problem have still been found~\cite{Herzog2020,Matsuki2021}.
In the present model, we find that, despite the modification, the profile of the Chern numbers around the modified 
spin-1 Dirac cones coincides with the conventional one.
 
The rest of this paper is structured as follows.
In Sec.~\ref{sec:model}, we introduce the model we 
study in this paper, namely, 
the spinless fermion model on a square kagome lattice.  
We closely analyze the band structures 
and elucidate two characteristic features, namely the flat band and the spin-1 Dirac cone.  
In Sec.~\ref{sec:Hof}, we show the Hofstadter diagram of the square kagome model,
paying particular attention to the topological gap-opening of the modified spin-1 Dirac cones.
Finally, we present the summary of this paper in Sec.~\ref{sec:summary}.
We remark that part of the results has already been shown in Ref.~\onlinecite{Kuno2020_scar}, 
where we discussed the possibility of the quantum scar state in this model in the presence of interactions.
Here we focus on a comprehensive understanding of 
the single-particle band structures and their topological properties, 
which has not been addressed in Ref.~\onlinecite{Kuno2020_scar}.

\begin{figure*}[tb]
\begin{center}
\includegraphics[clip, width = 0.95\linewidth]{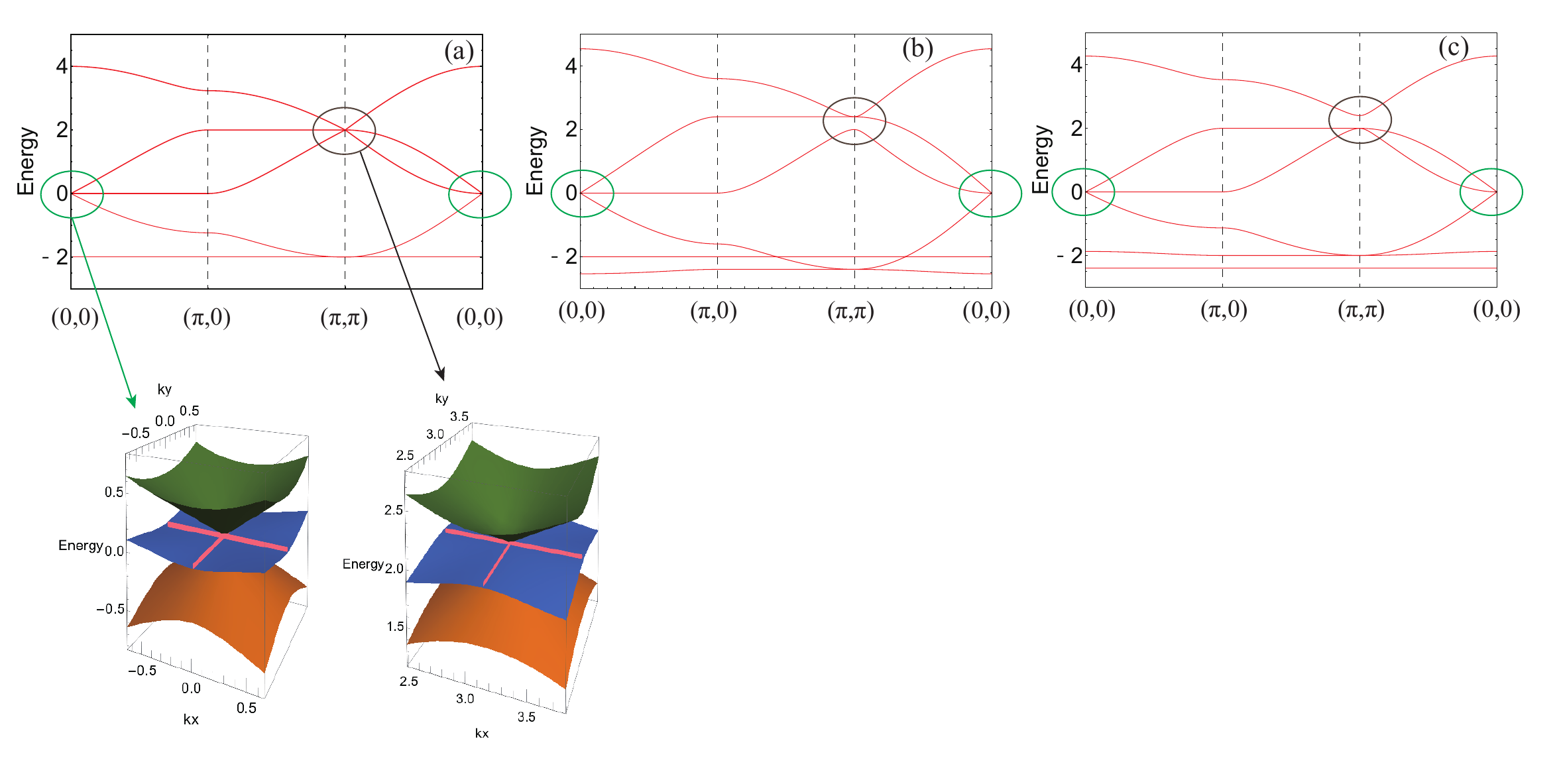}
\vspace{-10pt}
\caption{The band structures for the tight-binding model for 
(a) $t_1= 1$, $t_2 = 1$, (b) $t_1= 1$, $t_2 = 1.2$, and (c) $t_1= 1.2$, $t_2 =1$.
For (a), we show the zoom-up of the spin-1 Dirac cones. Pink lines correspond to the directions where the middle band has flat dispersion.}
 \label{fig:band}
 \end{center}
 \vspace{-10pt}
\end{figure*}

\section{Model and band structure \label{sec:model}}
We consider the tight-binding model without external field for spinless fermions
on a square kagome lattice with NN hoppings, 
$H = \sum_{\langle i,j \rangle} t_{i,j} c^{\dagger}_{i}c_j + (\mathrm{H.c.})$.
Here $t_{i,j}$ is the transfer integral between the sites $i$ and $j$, 
and $c_i$ ($c_i^\dagger$) stands for the annihilation (creation) operator at site $i$.
Each site is specified by the position of the unit cell $\bm{R} = R_x \bm{e}_x + R_y \bm{e}_y$ and the sublattice $\alpha  =1,\cdots 6$ (Fig.~\ref{fig:model}).

In the following, we consider the case where red (blue) bonds have transfer integral $t_1$ ($t_2$).
The Hamiltonian in the momentum-space representation is
\begin{eqnarray}
H = \sum_{\bm{k}} \bm{c}_{\bm{k}}^\dagger \mathcal{H}_{\bm{k}} \bm{c}_{\bm{k}}, \label{eq:Ham}
\end{eqnarray}
where $\bm{c}_{\bm{k}} = \left( c_{\bm{k},1}, \cdots,  c_{\bm{k},6} \right)^{\rm T}$, and 
\begin{eqnarray}
 \mathcal{H}_{\bm{k}} = 
 \begin{pmatrix}
 0 & t_1 & 0 & t_1 & t_2e^{-ik_x} &t_2e^{-ik_y}  \\
 t_1 & 0 & t_1 & 0 & t_2 & t_2 e^{-ik_y}  \\
 0&t_1 & 0&t_1 & t_2  & t_2 \\
 t_1 & 0 & t_1 & 0 & t_2 e^{-ik_x} & t_2 \\
 t_2 e^{ik_x} & t_2 & t_2 & t_2 e^{ik_x} & 0 & 0 \\
  t_2 e^{ik_y} & t_2 e^{ik_y}& t_2 & t_2 & 0 & 0 \\ 
 \end{pmatrix}. \label{eq:H_mom}
\end{eqnarray}
Diagonalizing $\mathcal{H}_{\bm{k}}$, we obtain the band structure. 
We plot the band structures for $t_1= t_2$ [Fig.~\ref{fig:band}(a)], 
$t_1< t_2$ [Fig.~\ref{fig:band}(b)], and $t_1> t_2$ [Fig.~\ref{fig:band}(c)] (here we set both $t_1$ and $t_2$ to be positive).
Clearly, several characteristics appear in the band structures, as we will discuss in detail below. 

\subsection{Flat bands \label{sec:FB}} 
We see in Fig.~\ref{fig:band} that a flat band with the energy $-2t_1$ exists.
For both $t_1$ and $t_2$ being positive, the flat band has the lowest energy for 
$t_1 \geq t_2$ whereas a dispersive band has the lowest energy for $t_1 < t_2$. 
Interestingly, for $t_1 < t_2$, the flat band intersects the 
dispersive band and their cross section forms a ring. 
Such a band structure was found in several models~\cite{Misumi2017,Mizoguchi2019_FBenginneer,Zheng2020} 
and was referred to as a type-III nodal ring~\cite{Zheng2020}. 

\begin{figure}[tb]
\begin{center}
\includegraphics[clip, width = 0.95\linewidth]{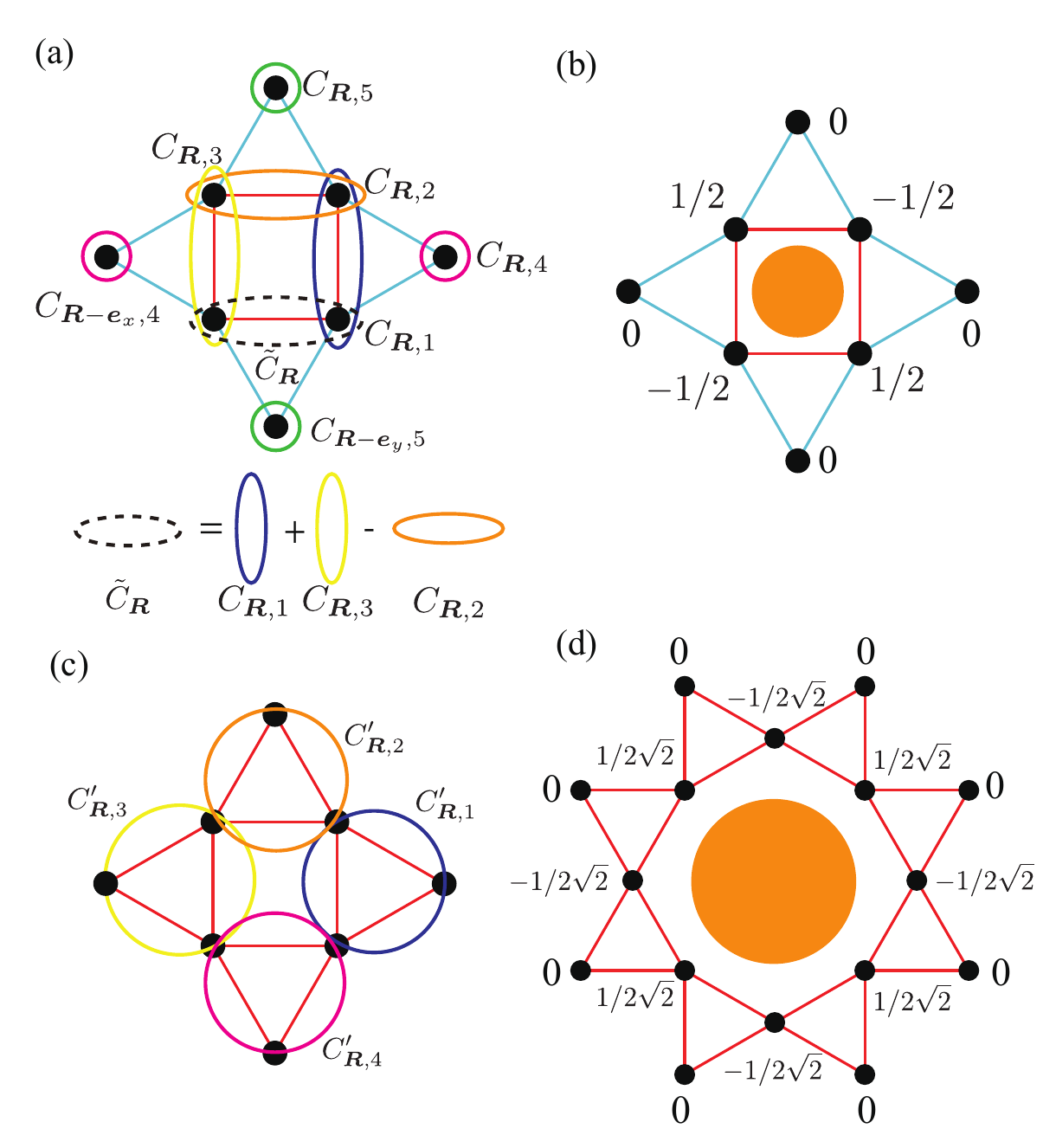}
\vspace{-10pt}
\caption{Schematic figure of 
(a) MOs and 
(b) a CLS on a square plaquette for any $t_1$ and $t_2$.
The panels (c) and (d) are MOs and an additional CLS on an octagonal plaquette, respectively, 
which exist only for $t_1 = t_2$.}
  \label{fig:MO_CLS}
 \end{center}
 \vspace{-10pt}
\end{figure}

The emergence of the flat band is accounted for by the MO representation~\cite{Hatsugai2011,Hatsugai2015,Mizoguchi2019,Mizoguchi2019_2,Mizoguchi2020}, 
which describes the generic flat-band models.
Specifically, we define the following fermion operators 
composed of the linear combination of the atomic orbitals:
\begin{subequations}
\begin{eqnarray}
C_{\bm{R},1} = c_{\bm{R},2} + c_{\bm{R},3}, 
\end{eqnarray}
\begin{eqnarray}
C_{\bm{R},2} = c_{\bm{R},3} + c_{\bm{R},4}, 
\end{eqnarray}
\begin{eqnarray}
C_{\bm{R},3} = c_{\bm{R},4} + c_{\bm{R},1}, 
\end{eqnarray}
\begin{eqnarray}
C_{\bm{R},4} = c_{\bm{R},5} , 
\end{eqnarray}
and 
\begin{eqnarray}
C_{\bm{R},5} = c_{\bm{R},6}. 
\end{eqnarray}
\end{subequations}
We additionally define
\begin{eqnarray}
\tilde{C}_{\bm{R}}  = c_{\bm{R},1} + c_{\bm{R},2},
\end{eqnarray}
which satisfies 
\begin{eqnarray}
\tilde{C}_{\bm{R}}  = C_{\bm{R},1}  + C_{\bm{R},3} -C_{\bm{R},2}.  \label{eq:tildeC}
\end{eqnarray}
We call $C$-operators the MOs.
For the schematics of the MOs, see Fig.~\ref{fig:MO_CLS}(a).

Using these MOs, the Hamiltonian of Eq.~(\ref{eq:Ham}) can be written as
\begin{eqnarray}
H &=&\sum_{\bm{R}} t_1 \left(  
C^\dagger_{\bm{R},1} C_{\bm{R},1}  + C^\dagger_{\bm{R},2} C_{\bm{R},2}  + C^\dagger_{\bm{R},3} C_{\bm{R},3}  + 
\tilde{C}^\dagger_{\bm{R}} \tilde{C}_{\bm{R}}  
\right) \nonumber \\
&+& t_2 \left(C^\dagger_{\bm{R},1} C_{\bm{R},4}  +
C^\dagger_{\bm{R},2} C_{\bm{R},5}  + C^\dagger_{\bm{R},3} C_{\bm{R} -\bm{e}_x ,4}  
+ \tilde{C}^\dagger_{\bm{R}} C_{\bm{R}-\bm{e}_y,5}  
\right) \nonumber \\
&+& (\mathrm{H.c.})\nonumber \\ 
&-& 2t_1\sum_{i} c_i^\dagger c_i.
\label{eq:Ham_MO_real}
\end{eqnarray}
Recalling that $\tilde{C}_{\bm{R}}$ is linearly dependent on other MOs, we find from (\ref{eq:Ham_MO_real}) that 
the Hamiltonian (up to the constant energy shift of $- 2t_1\sum_{i} c_i^\dagger c_i$) can be written by using five degrees of freedom per unit cell, whereas each unit cell contains six atomic sites.
This reduction of the number of degrees of freedom is the origin of the flat band. 
To be more concrete, performing the Fourier transformation and using Eq.~(\ref{eq:tildeC}), 
we find that the Hamiltonian matrix in the momentum space,  
$\mathcal{H}_{\bm{k}}$, can be written as
\begin{eqnarray}
\mathcal{H}_{\bm{k}} = \Psi_{\bm{k}} h_{\bm{k}}  \Psi^{\dagger}_{\bm{k}} - 2t_1 I_6,  \label{eq:MOrep}
\end{eqnarray}
where
\begin{eqnarray}
\Psi_{\bm{k}} =  
\begin{pmatrix}
0 & 0& 1 & 0 & 0\\
1 & 0&  0 &0 & 0\\
1 & 1&  0 &0 & 0\\
0 & 1&  1 &0 & 0\\
0& 0 &0 &1&0 \\
0 & 0&  0& 0  &1 \\
\end{pmatrix}
\end{eqnarray}
and 
\begin{widetext}
\begin{equation}
\label{eq:h}
h_{\bm{k}}
=\begin{pmatrix}
2 t_1 & -t_1 & t_1 & t_2 & t_2 e^{-ik_y} \\
-t_1 & 2t_1 & -t_1 & 0 & t_2 \left( 1 - e^{-ik_y} \right) \\
t_1 & -t_1 & 2t_1 & t_2 e^{-ik_x} & t_2e^{-ik_y} \\
t_2 & 0&t_2 e^{ik_x}& 2t_1&0 \\
t_2 e^{ik_y} &  t_2 \left( 1 - e^{ik_y} \right) & t_2e^{ik_y} & 0 & 2t_1 \\
\end{pmatrix}.
\end{equation}
\end{widetext}
Note that the $\ell$-th column of the $6 \times 5$ matrix $\Psi_{\bm{k}}$ corresponds to 
the momentum-space representation of the MOs $C_{\bm{R},\ell}$.
Namely, the relation,
\begin{eqnarray}
\bm{C}_{\bm{k}} = \Psi_{\bm{k}}^\dagger \bm{c}_{\bm{k}}
\end{eqnarray}
with $\bm{C}_{\bm{k}}  = (C_{\bm{k},1}, \cdots, C_{\bm{k},5})^{\rm T}$ holds.
We also note that the second term of Eq.~(\ref{eq:MOrep}) gives a mere constant shift of the energy and that
$\Psi_{\bm{k}}$ in this model is actually $\bm{k}$-independent.

Equation~(\ref{eq:MOrep}) indicates that the vector which belongs to the kernel of $\Psi_{\bm{k}}^\dagger$
becomes an eigenstate of $\mathcal{H}_{\bm{k}}$ with the 
eigenenergy $-2t_1$.
Actually, the kernel of $\Psi_{\bm{k}}^\dagger$ can easily be found, that is, 
\begin{eqnarray}
\bm{u}_{\bm{k}} = \frac{1}{2}
\begin{pmatrix}
1\\
-1\\
1\\
-1\\
0\\
0\\
\end{pmatrix}.
\end{eqnarray}
By performing the inverse Fourier transformation to $\bm{u}_{\bm{k}}$, 
we find that $H$ has localized eigenstates on every square plaquette [Fig.~\ref{fig:MO_CLS}(b)]:
\begin{eqnarray}
L_{\bm{R}}^\dagger 
&=& \frac{1}{\sqrt{N_{\rm{u.c.}}}} \sum_{\bm{k}} e^{i\bm{k}\cdot \bm{R}} \left(\bm{c}_{\bm{k}}^\dagger \cdot \bm{u}_{\bm{k}} \right) \nonumber \\
&=& \frac{1}{2} \left( c^\dagger_{\bm{R},1}- c^\dagger_{\bm{R},2} + c^\dagger_{\bm{R},3}-c^\dagger_{\bm{R},4} \right), 
\end{eqnarray}
which satisfies $[H,L^\dagger_{\bm{R}} ] = -2t_1 L^\dagger_{\bm{R}}$ and $\{C_{\bm{R},\ell},L^\dagger_{\bm{R}^\prime}\}= 0$. 
($N_{\rm{u.c.}}$ is the number of unit cells.)
Note that the symbols $[\cdot,\cdot]$ and $\{ \cdot,\cdot \}$ stand 
for the commutation and the anticommutation, respectively. 
We also note that the same wave function was found as a localized magnon 
mode in Ref.~\onlinecite{Derzhko2006}.
Such localized orbitals with finite support, 
corresponding to the real-space representation of the flat band eigenstates,
are found in many flat band models~\cite{Bergman2008,Huber2010} and 
are recently called the compact localized states (CLSs)~\cite{Flach2014,Santos2020,Kuno2020,Kuno2020_scar,Santos2020}.
Remarkably, $L^\dagger_{\bm{R}}$'s are orthogonal to each other, and 
the neighboring $L^\dagger_{\bm{R}}$'s are separated from each other by two NN bonds. 
This fact indicates that, the many-body state of $1/6$-filling $\ket{\Psi_{\rm L}} = \prod_{\bm{R}} L^\dagger_{\bm{R}} \ket{0}$, 
where $\ket{0}$ stands for the fermion vacuum, 
remains to be an eigenstate when introducing the NN interaction, $H_{\rm int} = V \sum_{\langle i,j \rangle} n_i n_j$,
where $n_i = c^\dagger_i c_i$,
because $H_{\rm int}\ket{\Psi_{\rm L}}  = 0$.
In fact, $\ket{\Psi_{\rm L}}$ has a Wigner-solid-like charge order~\cite{Wu2007}.
Further, as we have pointed out in prior work~\cite{Kuno2020_scar}, $\ket{\Psi_{\rm L}}$ 
becomes a quantum scar state 
if it does not have the lowest energy but is embedded in the middle of the many-body energy spectrum. 

We additionally note that the case of $t_1 = t_2$ is special
in that an alternative choice of the MOs is allowed, which leads to the two-fold degeneracy of the flat band.
To be specific, we use the following four MOs living on triangles, 
$C^\prime_{\bm{R},1}$-$C^\prime_{\bm{R},4}$, to rewrite the Hamiltonian.
Their explicit forms are given as
\begin{subequations}
\begin{eqnarray}
C^\prime_{\bm{R},1} = c_{\bm{R},2} + c_{\bm{R},3} + c_{\bm{R},5}, 
\end{eqnarray}
\begin{eqnarray}
C^\prime_{\bm{R},2} = c_{\bm{R},3} + c_{\bm{R},4} + c_{\bm{R},6}, 
\end{eqnarray}
\begin{eqnarray}
C^\prime_{\bm{R},3} = c_{\bm{R},1} + c_{\bm{R},4} + c_{\bm{R}-\bm{e}_x,5}, 
\end{eqnarray}
and 
\begin{eqnarray}
C^\prime_{\bm{R},4} = c_{\bm{R},2} + c_{\bm{R},1} + c_{\bm{R}-\bm{e}_y,6}.
\end{eqnarray}
\end{subequations}
See Fig.~\ref{fig:MO_CLS}(c) for their forms in real space.
Then the Hamiltonian for $t_1 = t_2$ can be rewritten as
\begin{eqnarray}
H &=& t_1 \sum_{\bm{R}} C^{\prime \dagger}_{\bm{R},1}C^\prime_{\bm{R},1} + C^{\prime \dagger}_{\bm{R},2}C^\prime_{\bm{R},2}
+C^{\prime \dagger}_{\bm{R},3} C^\prime_{\bm{R},3} + C^{\prime \dagger}_{\bm{R},4}C^\prime_{\bm{R},4} \nonumber \\
&-&2t_1 \sum_{i} c_i^\dagger c_i. 
\end{eqnarray}

\begin{figure}[b]
\begin{center}
\includegraphics[clip, width = 0.95\linewidth]{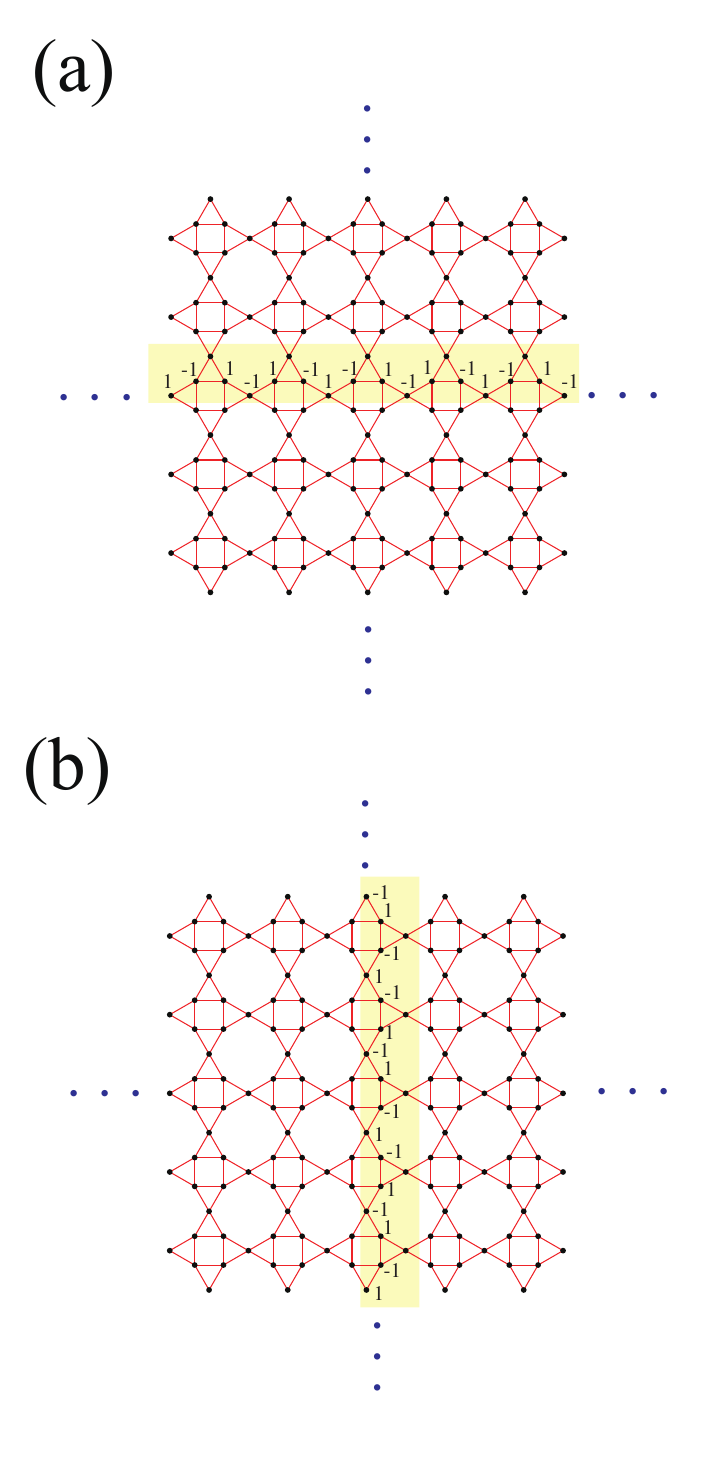}
\vspace{-10pt}
\caption{Schematics of the loop states. 
The loop is along (a) the $x$ direction and (b) the $y$ direction.
The numbers beside the sites denote the values of the wave function (up to the normalization constant).
The wave function is zero on the sites without the number.}
  \label{fig:loop}
 \end{center}
 \vspace{-10pt}
\end{figure}
In the momentum space, 
$\mathcal{H}_{\bm{k}}$ can be written as  
\begin{eqnarray}
\mathcal{H}_{\bm{k}} = \Psi^\prime_{\bm{k}} h^\prime_{\bm{k}}  \Psi^{\prime \dagger}_{\bm{k}} - 2t_1 I_6,  \label{eq:sqkagome_mo_2}
\end{eqnarray}
where
\begin{eqnarray}
\Psi^\prime_{\bm{k}} = 
\begin{pmatrix}
0 & 0& 1 & 1  \\
1 & 0&  0 &1  \\
1 & 1&  0 &0  \\
0 & 1&  1 &0  \\
1& 0 &e^{ik_x} &0 \\
0 & 1&  0& e^{ik_y}  \\
\end{pmatrix}
\end{eqnarray}
and 
\begin{eqnarray}
h^\prime_{\bm{k}} = t_1 I_4.
\end{eqnarray}
Again, the kernel of $\Psi^{\prime \dagger}_{\bm{k}}$ equals the eigenspace of $\mathcal{H}_{\bm{k}}$ with the eigenenergy $-2t_1$.
As the dimension of the kernel of $\Psi^{\prime \dagger}_{\bm{k}}$ is 2, the flat band for this parameter becomes doubly degenerate.
Furthermore, at $\bm{k} = (\pi, \pi)$, the dimension of the kernel of  $\Psi^{\prime \dagger}_{\bm{k}}$ becomes 3,
meaning that one additional mode having energy $-2t_1$ appears at this momentum. 
In fact, this leads to the quadratic band touching between the flat band and the dispersive band, 
which occurs in various flat-band models including the conventional kagome lattice~\cite{Bergman2008,Bilitewski2018,Rhim2019,Mizoguchi2019}.
\begin{figure}[tb]
\begin{center}
\includegraphics[clip, width = 0.95\linewidth]{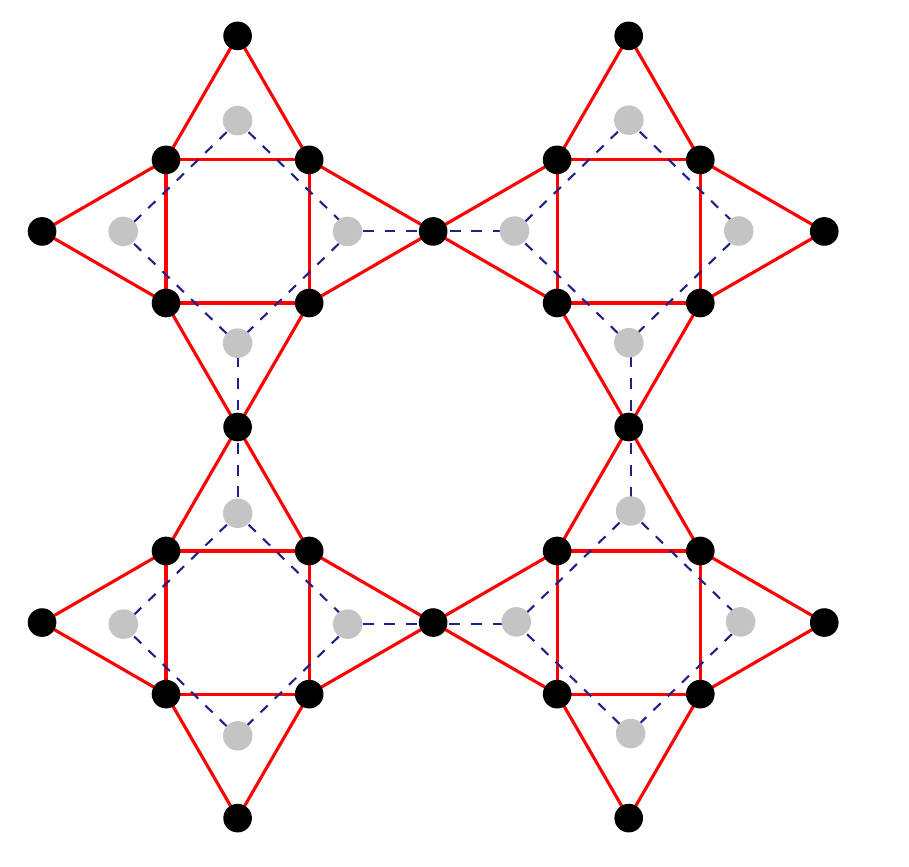}
\caption{Duality between the square kagome lattice and the square octagon lattice.
The sites on the square octagon lattice are depicted as gray dots.}
  \label{fig:octagon}
 \end{center}
 \vspace{-10pt}
\end{figure}

How are the CLSs changed in accordance with the change of the MOs?
To see this, 
we first note that $\{L_{\bm{R}}, C^{\prime \dagger}_{\bm{R}^\prime,\ell} \} = 0$ ($\ell = 1,2,3,4$),
thus $L_{\bm{R}}$ still serves as the CLS. 
In addition, the two-fold degeneracy of the flat band indicates the 
existence of additional species of the CLSs. 
In fact, such CLSs are located in each octagonal plaquette:
\begin{eqnarray}
L^{\prime \dagger}_{\bm{R}} &=& \frac{1}{2\sqrt{2}} \left[ c^\dagger_{\bm{R},3} + c^\dagger_{\bm{R} + \bm{e}_x,4} + c^\dagger_{\bm{R} + \bm{e}_x + \bm{e}_y,1} +  c^\dagger_{\bm{R} + \bm{e}_y,2}  \right] \nonumber \\
 &-& \frac{1}{2\sqrt{2}} \left[c^\dagger_{\bm{R},5} +c^\dagger_{\bm{R}+\bm{e}_x,6}+c^\dagger_{\bm{R} + \bm{e}_y ,5}+c^\dagger_{\bm{R},6} \right],
\end{eqnarray}
which satisfies $[H,L^{\prime \dagger}_{\bm{R}}] = -2t_1L^{\prime \dagger}_{\bm{R}}$ and 
$\{L^\prime_{\bm{R}}, C^{\prime \dagger}_{\bm{R}^\prime,\ell} \} = 0$.
See Fig.~\ref{fig:MO_CLS}(d) for its form in real space.
To the best of our knowledge, the CLS of Fig.~\ref{fig:MO_CLS}(d) has not been presented before.
Note that $L^\prime_{\bm{R}}$'s are not orthogonal to each other as the neighboring $L^\prime_{\bm{R}}$'s share a site. 
In addition,
$L_{\bm{R}}$'s and $L^\prime_{\bm{R}}$'s are not orthogonal to each other.
To be concrete, due to the fact that the two kinds of CLS share the sites of sublattices 1-4, 
they satisfy the following relation:
\begin{eqnarray}
\sum_{\bm{R}}  (-1)^{R_x + R_y} L^\dagger_{\bm{R}} 
= \sqrt{2} \sum_{\bm{R}} (-1)^{R_x + R_y} L^{\prime \dagger}_{\bm{R}}. 
\label{eq:LLp}
\end{eqnarray}
For the derivation of  Eq.~(\ref{eq:LLp}),  
we have used the following facts to evaluate the right-hand side, 
both of which arise from the sign factor $(-1)^{R_x + R_y}$.
(i) The atomic sites of sublattices 5 and 6 are vanishing after taking the summation over $\bm{R}$, and 
(ii) the atomic sites of sublattices 2 and 4 acquire the opposite sign factors to those of 1 and 3 after taking the summation over $\bm{R}$. 
From (i) and (ii), we find that the right-hand side of Eq.~(\ref{eq:LLp}) is equal to the left-hand side.

Equation (\ref{eq:LLp}) indicates that one of the CLSs is linearly dependent of the others, 
thus the number of linearly independent CLSs is $2N_{\rm u.c.}-1$.
Meanwhile, quadratic band touching at $(k_x,k_y) = (\pi, \pi)$ indicates that the number of 
states having the energy $-2t_1$ is $2N_{\rm u.c.}+1$,
which deviates from the number of linearly independent CLSs by $2$.
This can be compensated by the loop states (Fig.~\ref{fig:loop}) that wind the torus on which the system is placed 
(when the periodic boundary condition is imposed), as is the case of the kagome lattice~\cite{Bergman2008}.

\subsection{Duality to the square-octagon model}
Here we remark that the MO representation 
in the case of $t_1 = t_2$ is related to the duality between the square kagome lattice and the square octagon lattice. 
Namely, if we put a site at the center of every triangular plaquette (where the MOs are placed), we obtain the square octagon lattice (Fig.~\ref{fig:octagon}).  
In fact, due to this duality, the Hamiltonian matrix of the NN hopping model on a square octagon lattice,
which we denote $\mathcal{H}_{\bm{k}}^{\rm oct}$, is written as 
\begin{eqnarray}
\mathcal{H}_{\bm{k}}^{\rm oct} = t_1 \Psi_{\bm{k}}^{\prime \dagger} \Psi_{\bm{k}}^{\prime} -3 t_1 I_4. \label{eq:ham_oct}
\end{eqnarray}
In fact, it follows from Eqs.~(\ref{eq:sqkagome_mo_2}) and (\ref{eq:ham_oct}) that the dispersion relations 
of the square kagome model and that for the square octagon model are the same up to the constant shift,
because the eigenvalues of $\Psi_{\bm{k}}^{\prime \dagger} \Psi_{\bm{k}}^\prime$ are the same as 
those for $\Psi_{\bm{k}} \Psi^{\prime \dagger}_{\bm{k}}$ 
besides the two zero modes for $\Psi_{\bm{k}} \Psi^{\prime \dagger}_{\bm{k}}$~\cite{remark}.
Therefore, the spin-1 Dirac cones which we will discuss in the next section appear in the square octagon model as well~\cite{Kargarian2010}.
The square octagon lattice also attracts 
attention as a platform for exotic phenomena~\cite{Kargarian2010,Kariyado2014,Kudo2017,Pal2018,Lima2019,Nunes2020},
so we expect that this perspective of the MO representation will be useful for further research.  
 
\subsection{Spin-1 Dirac cones}
Another interesting feature of the square kagome fermion model is the emergence of the triple band touchings 
composed of two linearly dispersive bands and one quadratic band. 
For $t_1 = t_2$, they appear at $\bm{k} = (0,0)$ and $\bm{k} = (\pi, \pi)$, 
whose energies are $0$ and $2t_1$, respectively,
while that at $\bm{k} = (\pi, \pi)$ is gapped out when $t_1 \neq t_2$. 
Such band structures are referred to as the spin-1 Dirac cone 
and carries the monopole of charge 2~\cite{Berry1984}.
In tight-binding models, it appears in various systems such as  a Lieb lattice~\cite{Lieb1989} and related systems
~\cite{Sutherland1986,Shima1993,Vidal1998,Mizoguchi2019_2}. 
Note that the spin-1 Dirac cone does not necessarily appear in a pairwise manner, 
i.e., it is an exception of the Nielsen-Ninomiya theorem~\cite{Dagotto1986}.

The low-energy effective Hamiltonian of the conventional spin-1 Dirac cone is written as
\begin{eqnarray}
H^{\rm D}_{\bm{k}} = v \left(\delta \bm{k} \right)\cdot \bm{S},  \label{eq:Ham_D}
\end{eqnarray}
where $v$ is the velocity of the Dirac fermion,
$\delta \bm{k}$ is the momentum measured from the Dirac point, 
and $\bm{S} = \left(S_x, S_y, S_z \right)$ is the spin operator of $S=1$.
In this form of the effective model, one has the dispersion relation 
$E_{\bm{k}} = 0, \pm v|\delta \bm{k}|$,
which contains the completely flat band.
However, looking at the band structure of the square kagome model more closely, 
we find unique features of the spin-1 Dirac cones that can not be described by the 
effective model of Eq.~(\ref{eq:Ham_D}):
(i) The middle band is not completely flat, but it is flat only in particular directions, 
namely, $(k_x, 0)$ and $(0,k_y)$ for the Dirac cone at $(0,0)$ and $(k_x, \pi)$ and $(\pi,k_y)$ for that at $(\pi,\pi)$.
[see pink lines in Fig.~\ref{fig:band}(a)]. 
(ii) Away from the lines of $k_x = \pm k_y$, two dispersive bands 
do not have particle-hole symmetric dispersion with respect to the Dirac point. 
Remarkably, these two features are in sharp contrast to typical examples~\cite{Lieb1989,Sutherland1986,Shima1993,Vidal1998}, 
due to the following reason.
In the typical models of the spin-1 Dirac cones, such as the Lieb lattice model, 
the middle band is the completely flat band and the entire spectrum is particle-hole symmetric, 
since the model is chiral symmetric and the number of sublattices under the bipartition has an imbalance.
On the other hand, the square kagome model does not have the chiral symmetry, resulting in (i) and (ii). 
To reproduce these features, one may have to incorporate
symmetry-allowed corrections in addition to Eq.~(\ref{eq:Ham_D}).
(The modern viewpoint of the symmetry-based analysis on the multiple band crossing was 
presented in, e.g., Ref.~\onlinecite{Bradlyn2016}.)

For a deeper understanding of the spin-1 Dirac cones, 
we employ the MO representation for a certain direction in the momentum space.
This argument has been applied to construct the type-III Dirac cones~\cite{Mizoguchi2020_typeIII} and 
here we demonstrate that this is also useful to understand the spin-1 Dirac cones.
We emphasize that the MO representation discussed in the following is different from that in Sec.~\ref{sec:FB}.
Indeed, there is not real-space representation of MOs as the MO representation is applicable only in a part of the momentum space.

We first deal with the Dirac cone at $\bm{k}= (0,0)$ with $E = 0$.
As we have seen, the dispersion of the middle band is flat on the lines of $k_x = 0$ and $k_y = 0$.
Here we focus on the line of $k_y=0$. On this line, the Hamiltonian matrix $\mathcal{H}_{(k_x,0)}$ can be written by the MO representation as
\begin{eqnarray}
\mathcal{H}_{(k_x,0)} = \Phi_{(k_x,0)} \bar{h}_{(k_x,0)}   \Phi^\dagger_{(k_x,0)}, \label{eq:MO_direc_E0}
\end{eqnarray}
with 
\begin{eqnarray}
\Phi_{(k_x,0)} = 
\begin{pmatrix}
1 & 0 & 1  & 0 & 0 \\
0 & 1 & -1 & 0 & 0 \\
0 & 1 & 1 & 0 & 0 \\
1 & 0 & -1 & 0 & 0 \\
0 & 0 & 0 & e^{i k_x} & 1 \\
0 & 0 & 0 & 1 & 1 \\
\end{pmatrix}
\end{eqnarray}
and 
\begin{eqnarray}
\bar{h}_{(k_x,0)}  = 
\begin{pmatrix}
\frac{t_1}{2} & \frac{t_1}{2}  & 0  & t_2 & 0 \\
\frac{t_1}{2} & \frac{t_1}{2}  & 0 & 0& t_2 \\
0 & 0 & -\frac{t_1}{2}   & 0  & 0 \\
t_2 & 0& 0 & 0  & 0 \\
0 & t_2 & 0 & 0 & 0\\
\end{pmatrix}.
\end{eqnarray}
From Eq.~(\ref{eq:MO_direc_E0}) we find that 
the vector,
$\bm{u}_{(k_x,0)} = \frac{1}{2}(1,1,-1,-1,0,0)^{\rm T}$,
is the zero mode for $\mathcal{H}_{(k_x,0)}$
because $\bm{u}_{(k_x,0)}$ satisfies $\Phi^\dagger_{(k_x,0)} \bm{u}_{(k_x,0)} = 0$.
Further, at the Dirac point (i.e, $k_x = 0$), the fourth and the fifth columns of $\Phi_{(k_x,0)}$ become identical.
Hence, at this point, $\mathcal{H}_{(0,0)}$ can be written in a further simplified form as 
\begin{eqnarray}
\mathcal{H}_{(0,0)} = \Phi^\prime_{(0,0)} \bar{h}^\prime_{(0,0)}   \Phi^{\prime \dagger}_{(0,0)}, \label{eq:MO_direc_E0_2}
\end{eqnarray}
with 
\begin{eqnarray}
\Phi^\prime_{(0,0)} = 
\begin{pmatrix}
1 & 0 & 1  & 0 \\
0 & 1 & -1 & 0 \\
0 & 1 & 1 & 0  \\
1 & 0 & -1 & 0 \\
0 & 0 & 0 & 1  \\
0 & 0 & 0 & 1\\
\end{pmatrix}
\end{eqnarray}
and 
\begin{eqnarray}
\bar{h}^\prime_{(0,0)}  = 
\begin{pmatrix}
\frac{t_1}{2} & \frac{t_1}{2}  & 0  & t_2 \\
\frac{t_1}{2} & \frac{t_1}{2}  & 0 & t_2 \\
0 & 0 & -\frac{t_1}{2}   & 0 \\
t_2 & t_2 & 0 & 0 \\
\end{pmatrix}.
\end{eqnarray}
From Eq.~(\ref{eq:MO_direc_E0_2}), we find that there exists an additional zero mode
for $\mathcal{H}_{(0,0)}$, that is, $\bm{u}^\prime_{(0,0)} = \frac{1}{\sqrt{2}}(0,0,0,0,1,-1)^{\rm T}$,
because $\bm{u}^\prime_{(0,0)}$ satisfies $\Phi^{\prime \dagger}_{(0,0)} \bm{u}^\prime_{(0,0)}=0$.

Furthermore, there exists one more zero mode of $\mathcal{H}_{(0,0)}$, which has a form 
\begin{eqnarray}
\bm{u}^{\prime \prime}_{(0,0)} =& \frac{1}{\mathcal{N}}  \Phi^{\prime}_{(0,0)} \mathcal{O}^{\prime -1} \bm{v}_{(0,0)} \nonumber \\
=& \frac{1}{2} (1,-1,-1,1,0,0)^{\rm T},
\end{eqnarray}
where $\mathcal{N}$ is the normalization constant,
\begin{eqnarray}
 \mathcal{O}^{\prime} = \Phi^{\prime \dagger}_{(0,0)} \Phi^\prime_{(0,0)}  = \mathrm{diag}(2,2,4,2),
\end{eqnarray}
is so-called the overlap matrix~\cite{Hatsugai2011,Mizoguchi2019},
and 
$\bm{v}_{(0,0)} = (1,-1,0,0)^{\rm T}$ is the vector satisfying $\bar{h}^\prime_{(0,0)} \bm{v}_{(0,0)} = 0$.
$\bm{u}^{\prime \prime}_{(0,0)}$ is the zero mode of $\mathcal{H}_{(0,0)}$ since the following relation holds:
\begin{eqnarray}
\mathcal{H}_{(0,0)}\bm{u}^{\prime \prime}_{(0,0)} &=& (\Phi^\prime_{(0,0)} \bar{h}^\prime_{(0,0)}   \Phi^{\prime \dagger}_{(0,0)}) 
\left( \frac{1}{\mathcal{N}}  \Phi^{\prime}_{(0,0)} \mathcal{O}^{\prime -1} \bm{v}_{(0,0)}  \right) \nonumber \\
&=&\frac{1}{\mathcal{N}}  \Phi^\prime_{(0,0)} \bar{h}^\prime_{(0,0)} 
\mathcal{O}^\prime \mathcal{O}^{\prime -1} \bm{v}_{(0,0)} \nonumber \\
&=& \frac{1}{\mathcal{N}}  \Phi^\prime_{(0,0)} \bar{h}^\prime_{(0,0)}  \bm{v}_{(0,0)}  = 0.
\end{eqnarray}
To summarize, there exists one zero mode on the line of $k_y=0$,
and at $\bm{k} = (0,0)$, the zero mode has three-fold degeneracy, which is nothing but the spin-1 Dirac cone;
this holds for arbitrary $t_1$ and $t_2$.

Next, we deal with the Dirac cone at $\bm{k}= (\pi,\pi)$ with $E=2 t_2$.
As we have seen, the Dirac cone acquires a mass gap when $t_1 \neq t_2$; at the Dirac point,
two out of three bands degenerate, i.e., the quadratic band touching occurs, 
and one of them has a flat dispersion along the line of $k_x = \pi$ and $k_y =\pi$,
only in the case of $t_1 = t_2 $, the quadratic band touching turns into the spin-1 Dirac cone.
For this Dirac cone, we focus on the line of $k_y=\pi$.
Then, we can write the Hamiltonian in the form of the MO representation as 
\begin{eqnarray}
\mathcal{H}_{(k_x,\pi)} = \Xi_{(k_x,\pi)}\tilde{h}_{(k_x,\pi)}   \Xi^{\dagger}_{(k_x,\pi)} + 2t_2 I_6, \label{eq:MOrep_pi}
\end{eqnarray}
with
\begin{eqnarray}
\Xi_{(k_x,\pi)} = 
\begin{pmatrix}
1 & 0 & 1  & -\frac{1}{2} & \frac{e^{-ik_x}}{2} \\
0 & 1 & -1 &  -\frac{1}{2} & \frac{1}{2} \\
0 & 1 & 1 &  \frac{1}{2} & \frac{1}{2}  \\
1 & 0 & -1 &  \frac{1}{2} & \frac{e^{-ik_x}}{2} \\
0 & 0 & 0 & 0&-1  \\
0 & 0 & 0 & -1 & 0\\
\end{pmatrix}
\end{eqnarray}
and 
\begin{eqnarray}
\tilde{h}_{(k_x, \pi)}  = 
\begin{pmatrix}
\frac{t_1-t_2}{2} & \frac{t_1 + t_2 e^{-ik_x}}{2} & 0 & 0 & 0\\
\frac{t_1 + t_2 e^{ik_x}}{2} & \frac{t_1-t_2}{2}  & 0 & 0 & 0 \\
0 & 0 & -\frac{t_1 + t_2}{2} & 0 & 0 \\
0 & 0 & 0 & -2t_2 & 0 \\
0 & 0 &0 & 0 & -2 t_2 \\
\end{pmatrix}. \nonumber \\
\end{eqnarray}
From Eq.(\ref{eq:MOrep_pi}), we can find that there exists an eigenmode with the eigenenergy 
$2t_2$, that is, $\tilde{\bm{u}}_{(k_x, \pi)} =\frac{1}{2\sqrt{2}}(-1,-1,1,1,0,2)^{\rm T}$.
The vector $\tilde{\bm{u}}_{(k_x, \pi)}$ satisfies $\Xi^\dagger_{(k_x, \pi)} \tilde{\bm{u}}_{(k_x, \pi)}  = 0$.

At $k_x = \pi$, $\tilde{h}_{(\pi, \pi)}$ becomes
\begin{eqnarray}
\tilde{h}_{(\pi, \pi)}  = 
\begin{pmatrix}
\frac{t_1-t_2}{2} & \frac{t_1 - t_2 }{2} & 0 & 0 & 0\\
\frac{t_1 - t_2}{2} & \frac{t_1-t_2}{2}  & 0 & 0 & 0 \\
0 & 0 & -\frac{t_1 + t_2}{2} & 0 & 0 \\
0 & 0 & 0 & -2t_2 & 0 \\
0 & 0 &0 & 0 & -2 t_2 \\
\end{pmatrix}. \nonumber \\ \label{eq:tildeh}
\end{eqnarray}
Unlike the Dirac cone at $\bm{k}=(0,0)$, we cannot write down the 
Hamiltonian by using the $4\times 6$ and $4 \times 4$ matrices.
However, we find that $\tilde{h}_{(\pi, \pi)}$ has one zero mode, 
$\tilde{\bm{v}}_{(\pi,\pi)} = (1,-1,0,0,0)^{\rm T}$.
Therefore, 
$\mathcal{H}_{(\pi,\pi)}$ has one additional eigenmode with eigenenergy $2t_2$, that is 
\begin{eqnarray}
\tilde{\bm{u}}^\prime_{(\pi,\pi)} = \frac{1}{\mathcal{N}^\prime}  \Xi_{(\pi,\pi)} \tilde{\mathcal{O}}^{ -1} \tilde{\bm{v}}_{(\pi,\pi)},
\end{eqnarray}
where $\mathcal{N}^\prime$ is the normalization constant and 
\begin{eqnarray}
\tilde{\mathcal{O}} = \Xi^\dagger_{(\pi,\pi)} \Xi_{(\pi,\pi)}
= \begin{pmatrix}
2 & 0 & 0 & 0  & -1 \\
0 & 2 & 0 & 0 &1 \\
0 & 0 & 4 & 0 & 0 \\
0 & 0 & 0 & 2 &0 \\
-1&1& 0& 0 & 2 \\
\end{pmatrix}.
\end{eqnarray}
This is the origin of the quadratic band touching.
Further, only in the case of $t_1 = t_2$, we have 
\begin{eqnarray}
\tilde{h}_{(\pi, \pi)}  = 
\begin{pmatrix}
0&0 & 0 & 0 & 0\\
0 &0 & 0 & 0 & 0 \\
0 & 0 & -t_1& 0 & 0 \\
0 & 0 & 0 & -2t_1 & 0 \\
0 & 0 &0 & 0 & -2 t_1 \\
\end{pmatrix}, \nonumber \\ \label{eq:tildeh_2}
\end{eqnarray}
which indicates that there is an additional zero mode of $\tilde{h}_{(\pi, \pi)}$, that is, 
$\tilde{\bm{v}}^\prime_{(\pi,\pi)} = (1,1,0,0,0)^{\rm T}$.
Therefore, $\mathcal{H}_{(\pi,\pi)}$ also has one additional eigenmode with eigenenergy $2t_2$, 
that is, 
\begin{eqnarray}
\tilde{\bm{u}}^{\prime \prime}_{(\pi,\pi)}= \frac{1}{\mathcal{N}^{\prime \prime}}  \Xi_{(\pi,\pi)} \tilde{\mathcal{O}}^{ -1} \tilde{\bm{v}}^\prime_{(\pi,\pi)},
\end{eqnarray}
where $\mathcal{N}^{\prime \prime}$ is the normalization constant.
This is the origin of the massless spin-1 Dirac cone at $\bm{k} = (\pi, \pi)$ for $t_1 = t_2$.

\begin{figure}[tb]
\begin{center}
\includegraphics[clip, width = 0.95\linewidth]{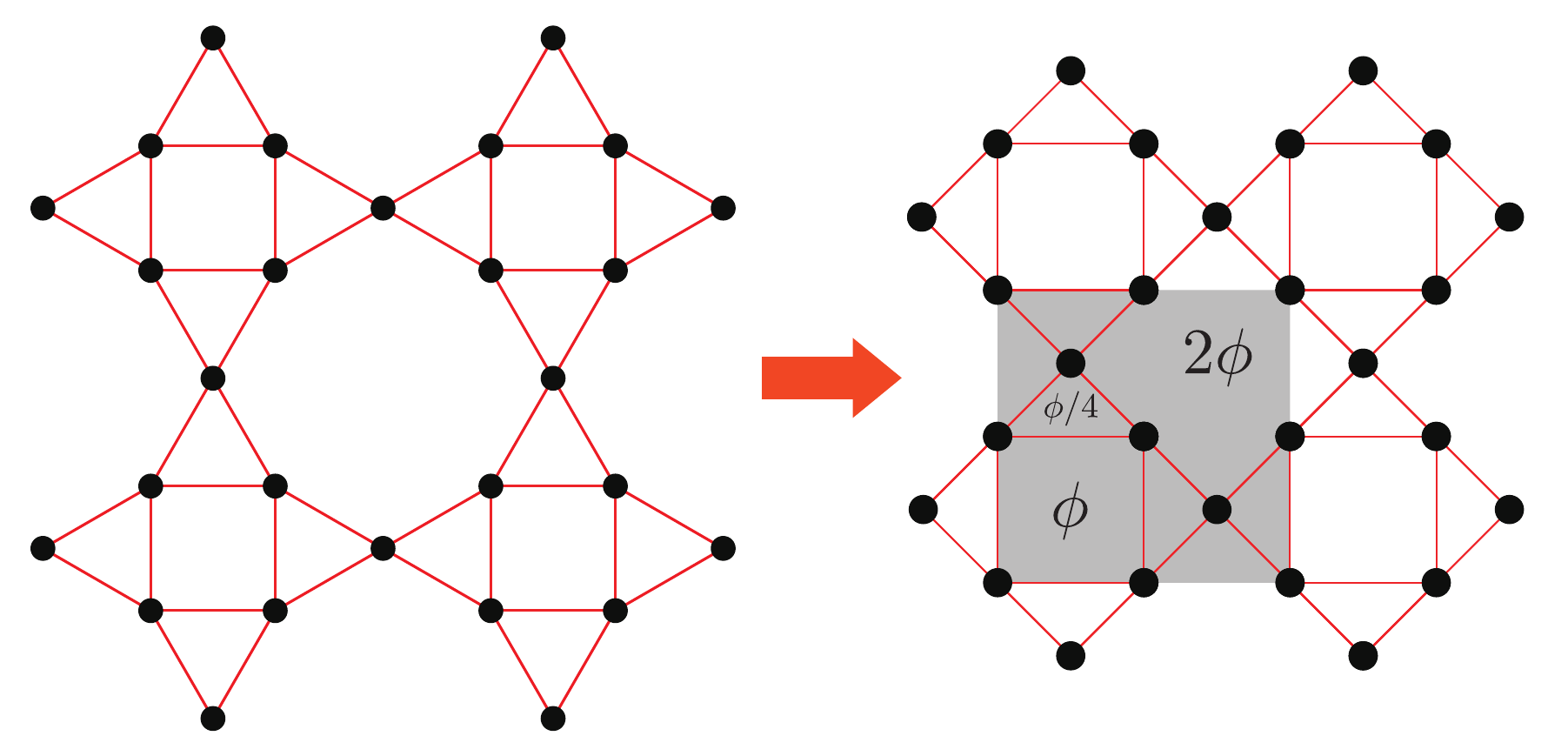}
\caption{Schematic of the flux distribution considered in this paper.
The gray shade denotes the area of the unit cell (in the absence of the magnetic flux).}
  \label{fig:flux}
 \end{center}
 \vspace{-10pt}
\end{figure}
\begin{figure*}[t]
\begin{center}
\includegraphics[clip, width = 0.9\linewidth]{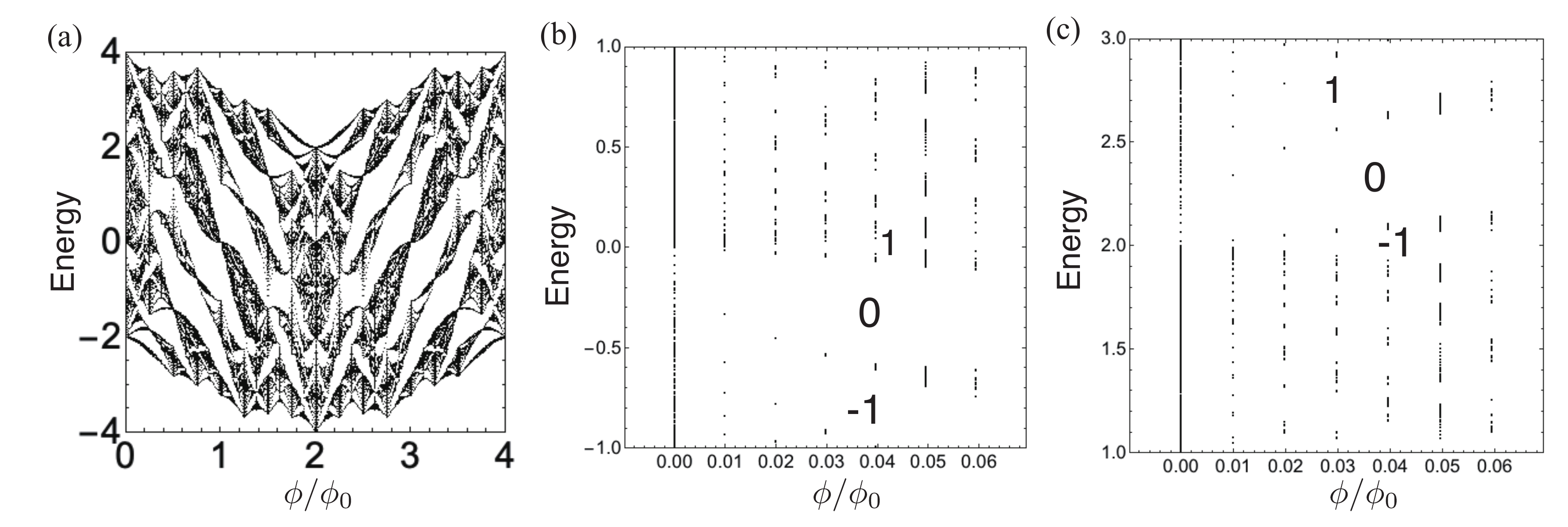}
\caption{(a) The Hofstadter diagram of the square kagome model.
The zoom-up in the low field for
(b) near $E=0$ and (c) $E = 2$.
The numbers at the blanks are the Chern numbers. }
  \label{fig:Hofstadter}
 \end{center}
 \vspace{-10pt}
\end{figure*}
 
\section{Hofstadter problem \label{sec:Hof}} 
As is well-known, 
topological gap-opening of Dirac cones in two dimensions leads to topological insulators 
with a finite Chern number~\cite{Semenoff1984,Haldane1988}.
Turning to the present model, the emergence of the modified spin-1 Dirac cones poses a question:
How does the Chern number behave upon topological gap opening in the modified spin-1 Dirac cones?
To address this question, we consider the Hofstadter problem on a square kagome lattice. 
For simplicity, we consider the case of $t_1 = t_2 = 1$.
We also deform the lattice 
with keeping the connectivity of the bonds of the square kagome lattice,
 as shown in Fig.~\ref{fig:flux},
 so that the flux penetrating each plaquette (shaped in either a triangule or a square) becomes rational. 
 
In Fig.~\ref{fig:Hofstadter}(a), we draw the Hofstadter diagram obtained by the numerical 
diagonalization on a finite system. 
The flux per unit cell is set to be $4\phi$, where $\phi$ is given as
$\phi = \frac{P}{Q} \phi_0$. 
Here, $Q$ is a prime number, $P= 0,\cdots 4Q$, and $\phi_0$ is 
a flux quantum. 
The system considered here contains $Q$ unit cells in $x$ direction and $N_{\rm mesh}$
unit cells in $y$ direction, so the total number of unit cells is $N_{\rm u.c.}= Q N_{\rm mesh}$.
We assign the periodic boundary condition in both $x$ and $y$ directions. 
To obtain Fig.~\ref{fig:Hofstadter}(a), we set $Q= 101$ and $N_{\rm mesh} = 24$.
We see a complex pattern of gap-opening, 
which somewhat resembles that for other frustrated lattices such as checkerboard (Mielke)~\cite{Aoki1996} and kagome lattices~\cite{Kimura2002}.

To investigate the topological gap opening of the spin-1 Dirac cones, 
we focus on the low-field limit and calculate the Chern number. 
To obtain the Chern number at a certain gap from the diagram, 
we employ the Streda formula~\cite{Streda1982}:
\begin{eqnarray}
\nu = \frac{\Delta (N_{\rm state}/N_{\rm u.c.}) }{4 \Delta (\phi/\phi_0)}, \label{eq:streda}
\end{eqnarray}
where $\nu$ stands for the total Chern number 
at the gap below which the states are occupied,
and $N_{\rm state}$ is the number of states below the gap.
In the denominator of Eq.~(\ref{eq:streda}), the interval $\Delta (\phi/\phi_0)$ is $1/Q$,
and the factor $4$ reflects the fact that the flux per unit cell is $4\phi$.
The Chern numbers for the gaps are shown in 
Figs.~\ref{fig:Hofstadter}(b) and \ref{fig:Hofstadter}(c) for the Dirac cone at $E=0$ 
and $E=2$, respectively. 
Here we use the data for $\phi/\phi_0 \in \left[\frac{1}{101} , \frac{6}{101} \right]$.
In both of two cases, the Chern number changes as $-1 \rightarrow 0 \rightarrow 1$.
Note that the change of the Chern number by 1, which is the half of the monopole charge, 
originates from the fact that in two-dimensions
the magnetic flux penetrating only one of hemispheres around the Dirac point can contribute to the Chern number~\cite{Watanabe2010,Watanabe2011}.
This behavior coincides with the case of the single massless spin-1 Dirac cone~\cite{Xu2017}, 
although the spectrum is highly particle-hole asymmetric.
Note that we do not see the field-insensitive energy modes, which are characteristic of the 
conventional spin-1 Dirac systems~\cite{Aoki1996}. This is attributed to the absence of the 
sublattice-imbalanced chiral symmetry, as pointed out before.

\section{Summary \label{sec:summary}}
We have investigated the band structures 
of the tight-binding model on a square kagome lattice. 
In the NN hopping model without external field, 
the flat band and the spin-1 Dirac cones appear. 
As for the flat band, we elucidate its origin by using the MO representation. 
As for the spin-1 Dirac cones, they appear in $(0,0)$ and $(\pi,\pi)$, each of which has different energy.
They also exhibit interesting features, namely the bending of the middle band and the particle-hole asymmetry, 
which indicates the modification term to the conventional Dirac Hamiltonian is necessary.
We further find that, in the presence of the external field, 
the topological band gap appears at the modified spin-1 Dirac cones,
and the Chern number of several gaps around the cones behaves in the same way as the conventional one. 

To conclude, we hope that our work provides a renewed view of the square kagome model as a playground for characteristic band structures 
induced by geometrical frustration.
Also, studying the correlation effects will be an interesting direction for the future study. 
In fact, we have already pointed out that the CLSs associated with the flat band lead to the quantum scar state at $1/6$-filling~\cite{Kuno2020_scar}.
Besides, the localized spin model on this lattice is considered to be a candidate of quantum spin liquid~\cite{Siddharthan2001,Tomczak2003,Richter2004,Derzhko2006,Wildeboer2011,Nakano2013,Derzhko2013,Rousochatzakis2013,Derzhko2014,Ralko2015,Pohle2016,Morita2018,Hasegawa2018,Lugan2019}. 
As the localized spin model is obtained as a low-energy effective Hamiltonian of the Hubbard model with the half-filled system of the spinful fermions,
and the spin-1 Dirac cone at $\bm{k}=(0,0)$ is at the Fermi energy in that case, 
the strongly correlated spin-1 Dirac fermions are expected to play a crucial role in the (possible) quantum spin liquid.
We expect that a variety of exotic phases will appear 
when we consider the various fillings or incorporate the spin degrees 
of freedom for the square kagome fermion model.
Studying the correlation effects on this model will be an intriguing future problem.  

\acknowledgements
T. M. thanks K. Kudo for fruitful discussions. 
This work is supported by the JSPS KAKENHI, 
Grants No.~JP17H06138 and No. JP20K14371 (T. M.), Japan.

\bibliographystyle{apsrev4-2}
\bibliography{sqkagome}

\begin{thebibliography}{70}
\bibitem [{rem()}]{remark}%
  \BibitemOpen
  \href@noop {} {}\bibinfo {note} {We note that Eq.~(\ref{eq:sqkagome_mo_2})
  can be rewritten as $\mathcal{H}_{\bm{k}} =t_1 \Psi^\prime_{\bm{k}}
  \Psi^{\prime \dagger}_{\bm{k}} - 2t_1 I_6$, as $h^\prime_{\bm{k}}$ is
  proportional to the identity matrix.}\BibitemShut {Stop}%
\end{thebibliography}
\end{document}